\documentclass[prd,aps,nofootinbib,onecolumn,showpacs,10pt]{revtex4}
\usepackage{graphicx,epsfig}

\def\pp{{\prime\prime}}
\def\vp{\varepsilon}

\begin{document}

\title{ $B_c$ to p-wave Charmonia Transitions in Covariant Light-Front Approach}
\author{ Xiao-Xia Wang,  Wei Wang,  and Cai-Dian L\"u}
\affiliation{
 \it  Institute of High Energy Physics \\
 \it and Theoretical Physics Center for Science Facilities,  Chinese Academy
of Sciences, Beijing 100049, People's Republic of China
 }

\begin{abstract}

In the covariant light-front quark model, we investigate the $B_c\to
h_c, \chi_{c0,1,2}$ form factors. The form factors are evaluated in
space-like kinematic region and are recasted to the physical region
by adopting the exponential parametrization.  We also study the
semileptonic $B_c$ decays and find that branching fractions for the
$B_c\to (h_c,\chi_{c0,1,2})l\bar\nu (l=e,\mu)$ decays have the order
$10^{-3}$ while branching fractions for $B_c\to
(h_c,\chi_{c0,1,2})\tau\bar\nu_\tau$ are suppressed by one order.
These predictions will be tested on the forthcoming hadron
colliders.

\end{abstract}


 \maketitle


\section{Introduction}

$B$ meson weak decays provide a golden place to extract magnitudes
and phases  of the Cabibbo-Kobayashi-Maskawa (CKM) matrix elements,
which can test the origins for CP violation in and beyond the
standard model (SM). Semileptonic and nonleptonic $B$ meson decays
have received extensive interests and achieved many great successes.
Experimentally, the two $B$ factories have accumulated more than
$10^9$ $B-\bar B$ events; measurements are becoming more and more
precise. On the theoretical side, apart from contributions
proportional to the form factors,  the so-called nonfactorizable
diagrams and some other radiative corrections are also taken into
account.  All of these are making $B$ physics suitable for search of
new particles and new phenomena (see Ref.~\cite{Buchalla:2008jp} for
a review).

Compared with $B$ decays, $B_c$ meson decays have received much less
experimental considerations.  The mass of a $B_c\bar B_c$ pair has
exceeded the threshold of $\Upsilon(4S)$ thus the $B_c$ meson can
not be generated on the two B factories. But $B_c$ meson decays have
a promising prospect on the forthcoming hadron colliders. The Large
Hadron Collider (LHC) experiment, which is scheduled to run in the
very near future, will produce plenty of $B_c$ events. The LHCb
collaboration has the desire to perform a comprehensive
investigation on $B_c$ meson decays. With more and more data
accumulated in the future, the study on $B_c$ mesons will be of
great interests: (1) $B_c$ contains two different heavy flavors, the
spectroscopy  may be different with the light meson or the  meson
with only one heavy quark. It serves as a different laboratory to
study the strong interactions. (2) $B_c$ meson can weakly decay via
the $b\to q$ transition like lighter $B_{u,d,s}$ mesons, but  the
dynamics is dramatically different. (3) Moreover, the charm quark
can also decay via weak interactions, where the $b$ quark acts as a
spectator. The CKM matrix element $|V_{cs}|\sim 1$ is much larger
than $|V_{cb}|\sim 0.04$ in $b$ quark decays. Decays of the charm
quark contribute much more to the decay width of the $B_c$ meson.
Although the phase space in $c\to d,s$ decays is much smaller than
that in $b\to c$ transitions, the former ones provide about $70\%$
contributions to the decay width of $B_c$. This results in a larger
decay width and a much smaller lifetime than the $B$ meson:
$\tau_{B_c}<\frac{1}{3}\tau_{B}$. (4) The two heavy $b$ and $\bar c$
quarks can annihilate and they provide  new kinds of weak decays
with sizable partial decay widths. The pure leptonic or radiative
leptonic decay can be used to extract the $B_c$ decay constant and
the CKM matrix element $V_{cb}$~\cite{Chang:1994jp,Chang:1998jp}.

Semileptonic $B_c$ decays are simpler than nonleptonic $B_c$ decays:
the leptonic part can be perturbatively evaluated leaving only
hadronic form factors unknown. In two-body nonleptonic $B_c$ decays,
most channels are also dominated by the $B_c$ transition form
factors. Thus the study of $B_c$ transition form factors is
essentially required. In the present work, we will use the
light-front quark model to analyze the form factors $B_c$ decays
into p-wave chamonia. This can be viewed as a continuation of our
previous work~\cite{Wang:2007sxa}. The light front QCD approach has
some unique features which are particularly suitable to describe a
hadronic bound state~\cite{Brodsky:1997de}. Based on this approach,
a light-front quark model with many advantages is
developed~\cite{Jaus:1989au,Jaus:1991cy,Cheng:1996if,Choi:2001hg,Jaus:1999zv}.
This model can provide a relativistic treatment of the movement of
the hadron. Is also gives a fully treatment of the hadron spin by
using the so-called Melosh rotation.  The light front wave functions
are expressed in terms of their fundamental quark and gluon degrees
of freedom. They are independent of the hadron momentum and thus are
explicitly Lorentz invariant.  In the covariant light-front quark
model,  the spurious contribution, which is dependent on the
{orientation} of the light-front, becomes irrelevant in  physical
observbles and that makes the light-front quark model more
selfconsistent.  This model has been successfully extended to
investigate the decay constants and form factors of the $s$-wave and
$p$-wave mesons~\cite{Cheng:2003sm,Cheng:2004yj,Lu:2007sg}.

Our paper is organized as follows. The formalism of the covariant
light-front quark model and numerical results for form factors are
presented in the next section. The decay rates of semi-leptonic
$B_c$ decays are  discussed in Section~\ref{sec:semileptonic}. Our
conclusions are given in Section~\ref{conc}.

\section{Form factors in the covariant light-front quark model}\label{sec:formalism}

$B_c\to S,A$ ($S,A$ denotes a scalar meson or an axial-vector meson,
respectively)  form factors are defined by
\begin{eqnarray}
       \langle S(P^\pp)|A_\mu|B_c(P^\prime)\rangle &=&
        \left(P_\mu-\frac{m_{B_c}^2-m_S^{2}}{q^2}q_\mu\right) F_1^{B_cS}(q^2)
        +\frac{m_{B_c}^2-m_S^{2}}{q^2}q_\mu F_0^{B_cS}(q^2) ,\label{eq:B TO_S} \\
      \langle A(P^\pp,\vp^{\pp*})|A_\mu|B_c(P^\prime)\rangle &=&
       -\frac{1}{ m_{B_c}-m_{A}}\,\epsilon_{\mu\nu\alpha \beta}\vp^{\pp*\nu}P^\alpha
    q^\beta  A^{B_cA}(q^2),\label{eq:B TO A}    \\
\ \ \ \
      \langle A(P^\pp,\vp^{\pp*})|V_\mu|
    B_c(P^\prime)\rangle &=& -i\Big\{
         (m_{B_c}-m_{A})\vp^{\pp*}_\mu V_1^{B_cA}(q^2)-\frac{\vp^{\pp*}\cdot P}
         { m_{B_c}-m_{A}}\,
         P_\mu V_2^{B_cA}(q^2)    \nonumber \\
    && -2m_{A}\,{\vp^{\pp*}\cdot P\over
    q^2}\,q_\mu\big[V_3^{B_cA}(q^2)-V_0^{B_cA}(q^2)\big]\Big\},\label{eq:B TO A1}
 \end{eqnarray}
where $P=P^\prime+P^{\prime\prime}$, $q=P^\prime-P^{\prime\prime}$
and the convention $\epsilon_{0123}=1$ is adopted.  To smear the
singularity at $q^2=0$, the relation $V_3^{B_cA}(0)=V_0^{B_cA}(0)$
is required, and
\begin{eqnarray}
 V_3^{B_cA}(q^2)&=&\frac{m_{B_c}-m_{A}}{ 2m_{A}}
 V_1^{B_cA}(q^2)-\frac{m_{B_c}+m_{A}}{2m_{A}}\,V_2^{B_cA}(q^2).\label{eq:relation}
\end{eqnarray}

Form factors of $B_c$ decays into a tensor meson are defined by
\begin{eqnarray}
 \langle T(P^\pp,\vp^{\pp*})|V_\mu|\bar B_c(P^\prime)\rangle
 &=&h(q^2)\epsilon_{\mu\nu\alpha\beta}\vp^{\pp*\nu\lambda}P_\lambda
 P^\alpha q^\beta,\nonumber\\
 \langle T(P^\pp,\vp^{\pp*})|A_\mu|\bar B_c(P^\prime)\rangle
 &=&-i\left\{k(q^2)\vp^{\pp*}_{\mu\nu}P^\nu +\vp^{\pp*}_{\alpha\beta}P^\alpha P^\beta[P_\mu b_+(q^2) +q_\mu b_-(q^2)]
 \right\},\label{eq:B TO T}
\end{eqnarray}
where the polarization tensor, which satisfies $\epsilon_{\mu\nu}
P^{\pp\nu}=0$, is symmetric and traceless.  The spin-2 polarization
tensors can be constructed  using spin-1 polarization vectors:
\begin{eqnarray}
 &&\tilde \epsilon_{\mu\nu}(p,\pm2)=
 \epsilon_\mu(\pm)\epsilon_\nu(\pm),\;\;\;\;
 \tilde \epsilon_{\mu\nu}(p,\pm1)=\frac{1}{\sqrt2}
 [\epsilon_{\mu}(\pm)\epsilon_\nu(0)+\epsilon_{\nu}(\pm)\epsilon_\mu(0)],\nonumber\\
 &&\tilde \epsilon_{\mu\nu}(p,0)=\frac{1}{\sqrt6}
 [\epsilon_{\mu}(+)\epsilon_\nu(-)+\epsilon_{\nu}(+)\epsilon_\mu(-)]
 +\sqrt{\frac{2}{3}}\epsilon_{\mu}(0)\epsilon_\nu(0),
\end{eqnarray}
where $\epsilon$ is the polarization vector for a vector meson. If
the recoiling meson is moving on the plus direction of the $z$ axis,
their explicit structures are chosen as
\begin{eqnarray}
\epsilon_\mu(0)&=&\frac{1}{m_T}(|\vec p_T|,0,0,E_T),\nonumber\\
\epsilon_\mu(\pm)&=&\frac{1}{\sqrt{2}}(0,\mp1,-i,0),
\end{eqnarray}
where the $E_T$ and $\vec{p_T}$ is the energy and the magnitude of
the momentum of the tensor meson in the $B_c$ rest frame,
respectively.  $m_T$ denotes the tensor meson's  mass. In $B_c$
meson decays, it is useful to define a new polarization vector
$\epsilon_T$ for the tensor meson
\begin{eqnarray}
  &&\epsilon_{T\mu}(h) =\tilde
  \epsilon_{\mu\nu}(p,h)p_{B_c}^\nu
\end{eqnarray}
which satisfies
\begin{eqnarray}
  && \epsilon_{T\mu}(\pm2)=0,\;\;\;
  \epsilon_{T\mu}(\pm1)=\frac{1}{\sqrt2}\epsilon(0)\cdot
  p_{B_c}\epsilon_\mu(\pm),\;\;\;
  \epsilon_{T\mu}(0)=\sqrt{\frac{2}{3}}\epsilon(0)\cdot
  p_{B_c}\epsilon_\mu(0).
\end{eqnarray}
The contraction is evaluated as $\epsilon(0)\cdot p_{B_c}=m_{B_c}
|\vec p_T|/m_T$. We can see that the new polarization vector plays a
similar role with the polarization vector for a vector meson,
regardless of the nontrivial factors $\frac{1}{\sqrt2}$ or
$\sqrt{\frac{2}{3}}$. In analogy with the $B_c\to A$ transition, one
can define the following form factors for convenience:
\begin{eqnarray}
 A^{B_cT}&=&-(m_{B_c}-m_T)h(q^2),\;\;\;\; V_1^{B_cT}=-\frac{k(q^2)}{m_{B_c}-m_T},\;\;\;
  V_2^{B_c T}= (m_{B_c}-m_T)b_+(q^2),\nonumber\\
 V_0^{B_cT}(q^2)&=&\frac{m_{B_c}-m_T}{2m_T} V_1^{B_cT}(q^2)-\frac{m_{B_c}+m_T}{2m_T}
 V_2^{B_c T}(q^2)-\frac{q^2}{2m_T}b_-(q^2),\label{eq:B TO T1}
\end{eqnarray}
where these form factors $A^{B_cT},V_{0,1,2}^{B_cT}$ have nonzero
mass dimensions.

We will work in the $q^+=0$ frame and employ the light-front
decomposition of the momentum $P^{\prime}=(P^{\prime -}, P^{\prime
+}, P^\prime_\bot)$, where $P^{\prime\pm}=P^{\prime0}\pm
P^{\prime3}$, so that $P^{\prime 2}=P^{\prime +}P^{\prime
-}-P^{\prime 2}_\bot$. The incoming (outgoing) meson have the
momentum of $P^{\prime}=p_1^{\prime}+p_2$ (
$P^{\pp}=p_1^{\pp}+p_2$) and the mass of $M^\prime$ $(M^\pp)$. The
quark and antiquark inside the incoming (outgoing) meson have the
mass $m_1^{\prime(\pp)}$ and $m_2$ and the momenta are denoted as
$p_1^{\prime(\pp)}$ and $p_2$ respectively. These momenta can be
expressed in terms of the internal variables $(x_i,
p_\bot^\prime)$ as:
 \begin{eqnarray}
 p_{1,2}^{\prime+}=x_{1,2} P^{\prime +},\qquad
 p^\prime_{1,2\bot}=x_{1,2} P^\prime_\bot\pm p^\prime_\bot,
 \end{eqnarray}
with $x_1+x_2=1$. Using these internal variables, one can define
some useful quantities for the incoming meson:
\begin{eqnarray}
 M^{\prime2}_0
          &=&(e^\prime_1+e_2)^2=\frac{p^{\prime2}_\bot+m_1^{\prime2}}
                {x_1}+\frac{p^{\prime2}_{\bot}+m_2^2}{x_2},\quad\quad
                \widetilde M^\prime_0=\sqrt{M_0^{\prime2}-(m^\prime_1-m_2)^2},
\nonumber\\
 e^{(\prime)}_i
          &=&\sqrt{m^{(\prime)2}_i+p^{\prime2}_\bot+p^{\prime2}_z},\quad\qquad
 p^\prime_z=\frac{x_2 M^\prime_0}{2}-\frac{m_2^2+p^{\prime2}_\bot}{2 x_2
 M^\prime_0},
 \end{eqnarray}
here $e_i$ can be interpreted as the energy of the quark or the
antiquark and $M_0^\prime$ can be viewed kinematic invariant mass
of the meson system. The definition of the internal quantities for
the outgoing meson is similar. To calculate the amplitude for the
transition form factor, we require the following Feynman rules for
the meson-quark-antiquark vertices ($i\Gamma^\prime_M$):
\begin{eqnarray}
 i\Gamma_P^\prime &=& H_P'\gamma_5,\nonumber\\
i\Gamma_S^\prime&=&
      H^\prime_S,
\nonumber\\
i\Gamma_{^3A}^\prime&=&iH^\prime_{^3A}[\gamma_\mu+\frac{1}{W^\prime_{^3A}}(p^\prime_1-p_2)_\mu]\gamma_5,
\nonumber\\
i\Gamma_{^1A}^\prime&=&iH^\prime_{^1A}[\frac{1}{W^\prime_{^1A}}(p^\prime_1-p_2)_\mu]\gamma_5,
\nonumber\\
i\Gamma_T^\prime&=&i\frac{1}{2}H^\prime_{T}[\gamma_\mu-\frac{1}{W^\prime_{V}}(p^\prime_1-p_2)_\mu](p^\prime_1-p_2)_\nu.
\end{eqnarray}
For the outgoing meson, one should use $i(\gamma_0
\Gamma^{\prime\dagger}_M\gamma_0)$ for the corresponding vertices.

\begin{figure}
\includegraphics[scale=0.5]{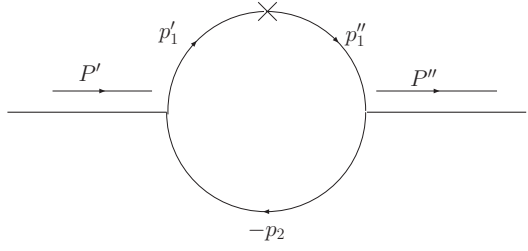}
\caption{Feynman diagram for $B_c\to S,A,T$ decay amplitudes. The X
in the diagram denotes the $V,A$ transition vertex while the
meson-quark-antiquark vertices are given in the
text.}\label{fig:feyn}
\end{figure}

In the conventional light-front quark model, the constituent quarks
are required to be on mass shell and the physical quantities can be
extracted from the plus component of the corresponding current
matrix elements. However, this framework suffers the problem of
non-covariance  because of the missing zero-mode contributions. In
order to solve this problem, Jaus proposed the covariant light-front
approach which permits a systematical way to deal with the zero-mode
contributions~\cite{Jaus:1999zv}. Physical quantities such as decay
constants and form factors can be calculated in terms of Feynman
momentum loop integrals which are manifestly covariant. The lowest
order contribution to a form factor is depicted in
Fig.~\ref{fig:feyn}. The X in the diagram denotes the $V,A$
transition vertex. In $B_c$ to p-wave chomonia decays,
$p_1'$($p_1''$) is the momentum of the bottom (charm) quark, while
$p_2$ is the momentum of the antiquark. It is also similar for the
notation of the quark masses.  As an example, we will derive the
$B_c\to S$ transition amplitude
\begin{eqnarray}
 {\cal B}^{B_cS}=-i^3\frac{N_c}{(2\pi)^4}\int d^4 p^\prime_1
 \frac{H^\prime_P ( H^\pp_S)}{N_1^\prime N_1^\pp N_2} S_\mu^{B_cS},
\end{eqnarray}
where $N_1^{\prime(\prime\prime)}= p_1^{\prime(\prime\prime)2}
-m_1^{\prime (\prime\prime)2} $, $N_2=p_2^2-m_2^2$. The function
$S^{B_cS}_\mu$ is derived using the Lorentz contraction
\begin{eqnarray}
 S^{B_cS}_\mu &=&{\rm Tr}\left[(\not \!p^\pp_1+m_1^\pp)\gamma_\mu\gamma_5(\not
 \!p^\prime_1+m_1^\prime)\gamma_5(-\not\!p_2+m_2)\right]\nonumber\\
           &=&2 p^\prime_{1\mu} [M^{\prime 2}+M^{\pp2}-q^2-2
           N_2-(m_1^\prime-m_2)^2-(m^\pp_1+m_2)^2+(m_1^\prime+m_1^\pp)^2]
 \nonumber\\
          &&+q_\mu[q^2-2 M^{\prime2}+N^\prime_1-N^\pp_1+2
         N_2+2(m_1^\prime-m_2)^2-(m_1^\prime+m_1^\pp)^2]
 \nonumber\\
          &&+P_\mu[q^2-N^\prime_1-N^\pp_1-(m_1^\prime+m_1^\pp)^2].
 \label{eq:SPPVappen}
\end{eqnarray}

In practice, we use the light-front decomposition of the loop
momentum and perform the integration over the minus component using
the contour method. If the covariant vertex functions are not
singular when performing the integration, the transition amplitude
will pick up the singularities in the antiquark propagator.  The
integration then leads to
 \begin{eqnarray}
 N_1^{\prime(\pp)}
      &\to&\hat N_1^{\prime(\pp)}=x_1(M^{\prime(\pp)2}-M_0^{\prime(\pp)2}),
 \nonumber\\
 H^{\prime(\pp)}_M
      &\to& h^{\prime(\pp)}_M,
 \nonumber\\
 W^\pp_M
      &\to& w^\pp_M,
 \nonumber\\
\int \frac{d^4p_1^\prime}{N^\prime_1 N^\pp_1 N_2}H^\prime_P H^\pp_S
S^{B_cS}
      &\to& -i \pi \int \frac{d x_2 d^2p^\prime_\bot}
                             {x_2\hat N^\prime_1
                             \hat N^\pp_1} h^\prime_P h^\pp_S \hat S^{B_cS},
 \end{eqnarray}
where
 \begin{eqnarray}
 M^{\pp2}_0
          =\frac{p^{\pp2}_\bot+m_1^{\pp2}}
                {x_1}+\frac{p^{\pp2}_{\bot}+m_2^2}{x_2}
 \end{eqnarray}
with $p^\pp_\bot=p^\prime_\bot-x_2\,q_\bot$. The explicit forms of
$h^\prime_M$ and $w^\prime_M$ used in this work are given by
\begin{eqnarray}
 h^\prime_P&=&(M^{\prime2}-M_0^{\prime2})\sqrt{\frac{x_1 x_2}{N_c}}
                    \frac{1}{\sqrt{2}\widetilde M^\prime_0}\varphi^\prime,\label{eqs:s-wave-wa}\\
 h^\prime_S&=&\sqrt{\frac{2}{3}}h^\prime_{^3A}
                  =(M^{\prime2}-M_0^{\prime2})\sqrt{\frac{x_1 x_2}{N_c}}
                    \frac{1}{\sqrt{2}\widetilde M^\prime_0}\frac{\widetilde
                    M^{\prime2}_0}{2\sqrt{3}M^\prime_0}\varphi^\prime_p,\\
h^\prime_{^1A}&=&h^\prime_{T}
                  =(M^{\prime2}-M_0^{\prime2})\sqrt{\frac{x_1 x_2}{N_c}}
                    \frac{1}{\sqrt{2}\widetilde M^\prime_0}\varphi^\prime_p
 \nonumber\\
                     w^\prime_{^3A}&=&\frac{\widetilde
                     M^{\prime2}_0}{m^\prime_1-m_2},\;\;
                    w^\prime_{^1A} =2\label{eqs:wa}
\end{eqnarray}
where $\varphi'$ and $\varphi'_p$ is the light-front wave function
for s-wave and p-pave mesons, respectively.  After this integration,
the conventional light-front model is recovered but manifestly the
covariance is lost as it receives additional spurious contributions
proportional to the lightlike four vector $\tilde\omega=(0,2,{\bf
0_{\perp}})$. The undesired spurious contributions can be eliminated
by the inclusion of the zero mode contribution which amounts to
performing the $p^-$ integration in a proper way in this approach.
The specific rules under this $p^-$ integration are derived in
Ref.~\cite{Jaus:1999zv,Cheng:2003sm} and are displayed in Appendix
A.

Using Eqs.~(\ref{eq:SPPVappen})--(\ref{eqs:wa}){ and taking the
advantage of the rules in Ref.~\cite{Jaus:1999zv,Cheng:2003sm}}, the
$B_c\to S$ form factors are straightforwardly given by
\begin{eqnarray}
 F_1^{B_cS}(q^2)&=&\frac{N_c}{16\pi^3}\int dx_2 d^2p^\prime_\bot
            \frac{h^\prime_P h^\pp_S}{x_2 \hat N_1^\prime \hat N^\pp_1}
            \bigg[x_1 (M_0^{\prime2}+M_0^{\pp2})+x_2 q^2
 \nonumber\\
         &&\qquad-x_2(m_1^\prime+m_1^\pp)^2 -x_1(m_1^\prime-m_2)^2-x_1(m_1^\pp+m_2)^2\bigg],
 \nonumber\\
 F_0^{B_cS}(q^2)&=& F_1^{B_cS}(q^2)+\frac{q^2}{q\cdot P}\frac{N_c}{16\pi^3}\int dx_2 d^2p^\prime_\bot
            \frac{2h^\prime_P h^\pp_S}{x_2 \hat N_1^\prime \hat N^\pp_1}
            \Bigg\{- x_1 x_2 M^{\prime2}-p_\bot^{\prime2}-m_1^\prime m_2
                  -(m_1^\pp+m_2)(x_2 m_1^\prime+x_1 m_2)
\nonumber\\
         &&\qquad +2\frac{q\cdot P}{q^2}\left(p^{\prime2}_\bot+2\frac{(p^\prime_\bot\cdot q_\bot)^2}{q^2}\right)
                  +2\frac{(p^\prime_\bot\cdot q_\bot)^2}{q^2}
                  -\frac{p^\prime_\bot\cdot q_\bot}{q^2}
                  \Big[M^{\pp2}-x_2(q^2+q\cdot P)
\nonumber\\
         &&\qquad -(x_2-x_1) M^{\prime2}+2 x_1 M_0^{\prime
                  2}-2(m_1^\prime-m_2)(m_1^\prime-m_1^\pp)\Big]
           \Bigg\}.
 \label{eq:fpm}
\end{eqnarray}
Similarly, one can derive the $B_c\to A, T$ form factors and we
refer to Appendix B for tedious expressions of these form factors.

The light front wave function $\varphi'$ can be obtained by solving
the relativistic Schr\"odinger equation with a phenomenological
potential. But in fact except for some limited cases, the exact
solution is not obtainable. In practice,   a phenomenological wave
function to describe the hadronic structure is preferred. In this
work, we will use the simple Gaussian-type wave function which has
been extensively examined in the
literature~~\cite{Cheng:2003sm,Cheng:2004yj,Lu:2007sg}
\begin{eqnarray}
 \varphi^\prime
    &=&\varphi^\prime(x_2,p^\prime_\perp)
             =4 \left({\pi\over{\beta^{\prime2}}}\right)^{3/4}
               \sqrt{{dp^\prime_z\over{dx_2}}}~{\rm exp}
               \left(-{p^{\prime2}_z+p^{\prime2}_\bot\over{2 \beta^{\prime2}}}\right),
\nonumber\\
 \varphi^\prime_p
    &=&\varphi^\prime_p(x_2,p^\prime_\perp)=\sqrt{2\over{\beta^{\prime2}}}
    ~\varphi^\prime,\quad\qquad
         \frac{dp^\prime_z}{dx_2}=\frac{e^\prime_1 e_2}{x_1 x_2
         M^\prime_0}.
 \label{eq:wavefn}
\end{eqnarray}

The parameters $\beta'$s, which describe the momentum distribution,
are usually fixed by mesons' decay constants whose analytic
expressions are also given in \cite{Cheng:2003sm}. The decay
constant for the $B_c$ meson is employed by
\begin{eqnarray}
 f_{B_c}=(400\pm40) {\rm MeV},
\end{eqnarray}
which gives $\beta_{B_c}=(0.89^{+0.075}_{-0.074})$ GeV.  This value
is a bit smaller than results provided by Lattice QCD
method~\cite{Chiu:2007km}
\begin{eqnarray}
f_{B_c}=(489\pm4\pm3) {\rm MeV}.
\end{eqnarray}
The other inputs, including  masses (in units of GeV) of the
constituent quarks and hadrons,  $V_{cb}$ and the lifetime of $B_c$,
are used as~\cite{Amsler:2008zz}
\begin{eqnarray}
 && m_c=1.4, \;\;\; m_b=4.8,\;\;\; m_{B_c}=6.286,\nonumber\\
 &&m_{h_c}=3.52528,\;\;\; m_{\chi_{c0}}=3.41476,\;\;\;  m_{\chi_{c1}}=3.51066,\;\;\;  m_{\chi_{c2}}=3.5562\nonumber\\
 &&V_{cb}=41.2\times 10^{-3},\;\;\; \tau_{B_c}=(0.46\pm0.07) ps.
\end{eqnarray}
The constituent quark masses are close to those used in the
literature~\cite{Cheng:2003sm,Cheng:2004yj,Lu:2007sg,Wang:2007sxa}.
The shape parameter for $\chi_{c1}$ is used as:
$\beta_{\chi_{c1}}=(0.7\pm0.1)$ GeV which corresponds to
$|f_{\chi_{c1}}|=(340^{+119}_{-101})$ MeV. For the other shape
parameters, we will assume the same values and introduce a
relatively large uncertainty to compensate the different Lorentz
structures:
$\beta_{\chi_{c0}}=\beta_{\chi_{c2}}=\beta_{h_c}=(0.7\pm0.1)$ GeV.

Unlike the light quark, the heavy bottom and charm quark have large
masses. In the heavy quark limit $m_{c,b}\to \infty$, the $B_c$ and
charmonium systems obey the heavy quark symmetry which is helpful to
simplify the dynamics in transition form factors and decay
amplitudes. In particular, the large momentum of the heavy quark can
be projected out and the remanent momentum is of the order of the
hadronic scale. For $B_c$ and charmonia, the two constituents are
both heavy and move non-relativistically. After projecting the large
mass scale, the dynamic scale  is of the order $m_cv$ and $m_cv^2$,
where $v$ is the relative velocity of the quark-anti-quark pair.
Then physical quantities can be expanded in terms of $1/m_c,1/m_b$
in the effective theory. In the present analysis, the expansion in
$1/m_c$ and $1/m_b$ is not used and physical quantities contain a
tower of contributions with different orders. The $B_c$ and the
charmonia are directly made of two heavy quarks and the dynamics is
reflected by the light-front wave functions. Despite of the
different treatments, the leading power behavior should be the same.
In $B_c$ and charmonia,  the transverse momentum is of the order of
$m_c v$. Inferred from Eq.~(\ref{eq:wavefn}), the parameter $\beta'$
in the light-front wave function is of the order of $m_c v$. In
nonrelativistic QCD (NRQCD)~\cite{Bodwin:1994jh}, the kinematic
energy has the order of $m_cv^2$. If it can be identified as the
hadronic scale , the $\beta'$ is of the order of $\sqrt {m_c
\Lambda_{\rm QCD}}$, which is expected to be larger than the shape
parameter in the light meson system. This feature is also confirmed
by the numerical result: the shape parameters $\beta'$s for
charmonium and $B_c$ meson are larger than those for light mesons
such as $\beta'_\pi=0.3102$ GeV~\cite{Cheng:2003sm}.

In the NRQCD framework,  the light-cone distribution amplitudes of
the s-wave charmonia have been comprehensively investigated in
Refs.\cite{Ma:2006hc,Braguta:2006wr,Braguta:2007fh,Bell:2008er}.
Recently, the analysis has been generalized to the p-wave
charmonia~\cite{Braguta:2008qe}. In their notation, the distribution
is described by the matrix elements of the nonlocal operators, while
in the light-front quark model, we use the coupling vertex.
Moreover, the distribution amplitudes in these two frameworks are
also different. In Ref.~\cite{Braguta:2008qe}, the leading twist
light-cone distribution amplitude is expanded into Gegenbauer
polynomials and the Gegenbauer moments are studied in the QCD sum
rules. The authors in  Ref.~\cite{Braguta:2008qe} also propose the
following model
\begin{eqnarray}
 \Phi(\xi,\mu\sim m_c) =c(\beta_P) (1-\xi^2)\xi {\rm
 exp}\left(-\frac{\beta_P}{1-\xi^2}\right),\label{eq:distributionamplitudes-sum-rule}
\end{eqnarray}
where $c(\beta_P)$ is a normalization constant. The parameter $\xi$
is defined as $\xi=2u-1$, where $u$ is the momentum fraction of the
charm quark. This form is dramatically different with the one in
Eq.~(\ref{eq:wavefn}) used in this framework. The main part in
Eq.~(\ref{eq:distributionamplitudes-sum-rule}) is the exponential
term $ {\rm
 exp}\left(-\frac{\beta_P}{1-\xi^2}\right)$ which corresponds to ${\rm exp}
\left(-{p^{\prime2}_z\over{2 \beta^{\prime2}}}\right)$ in the
distribution amplitude in Eq.~(\ref{eq:wavefn}). For charmonia, the
momentum $p_z^{\prime2}$ is simplified as
\begin{eqnarray}
p_z^{\prime2}=\frac{(x_1-x_2)^2}{x_1x_2}
(p^{\prime2}_{\perp}+m_c^2).
\end{eqnarray}
Considering the longitudinal part, we can see that there is an
additional factor $(x_1-x_2)^2=\xi^2$ in the one used in this
framework. It indicates that the distribution amplitude used in this
work is sharper at the region around $x_2\sim 0.5$.  At last, the
parameter $\beta_P$ is dimensionless and is different with $\beta'$.
In the heavy quark limit, $u$ is close to $1/2$ and $\beta_P$ is of
the order 1: $\beta_P=3.4^{+1.5}_{-0.9}$.Comparing the longitudinal
part of the two distribution amplitudes, one can obtain the relation
between $\beta'$ and $\beta_P$: $\beta_P\sim \frac{2m_c^2 \langle
(x_1-x_2)^2\rangle}{\beta'^2}$. The typical value of  $\langle
(x_1-x_2)^2\rangle$ is of the order $\frac{\Lambda_{\rm QCD}}{m_c}$
which also indicates $\beta_P\sim 1$. These different distribution
amplitudes are expected to induce sizable differences to the
resultant form factors. They can be discriminated or constrained by
the available data of transition form factors in the future.

Because of the condition $q^+=0$ imposed during the course of
calculation, form factors can be directly studied only at spacelike
momentum transfer $q^2=-q^2_\bot\leq 0$ which are not relevant for
the physical decay processes. It has been proposed in
\cite{Cheng:2003sm} to parameterize form factors as explicit
functions of $q^2$ in the space-like region and then analytically
extend them to the time-like region. To shed light on the momentum
dependence, one needs a specific model to parameterize the form
factors and we will choose a three-parameter form
\begin{eqnarray}
 F(q^2)=F(0){\rm exp}(c_1\hat s+c_2\hat s^2),
\end{eqnarray}
where $\hat s=q^2/m_{B_c}^2$ and $F$ denotes any one of the $B_c\to
S,A,T$ form factors. In the fitting procedure, form factors in the
region $q^2=[-1,-15]$ GeV$^2$ are studied  and the fitted results
for $c_1$ and $c_2$ are collected in
table~\ref{tab:resultsformfactors1}. We should point out that we
have adopted a negative decay constant for $\chi_{c1}$ so that the
$B_c\to A$ form factors are both positive.

\begin{table}
\caption{Results for the $B_c\to \chi_{c0,c1,c2},h_c$ form factors
and fitted parameters $c_1$ and $c_2$. The first type of
uncertainties are from the shape parameters of the p-wave charmonia
and the second ones are from the $B_c$ meson decay constants.
}\label{tab:resultsformfactors1}
\begin{tabular}{ccccccccccc}
 \hline\hline
 & $F$ & $F(0)$ &$F(q^2_{\rm {max}})$ & $c_1$ & $c_2$ \\
 \hline
 & $F_1^{B_c \chi_{c0}}$ & $ 0.47_{-0.06-0.01}^{+0.03+0.00}$  & $0.73_{-0.05-0.03}^{+0.01+0.02}$   & $ 2.03_{-0.24-0.10}^{+0.27+0.10}$  & $ 0.43_{-0.06-0.01}^{+0.05+0.01}$ \\
 &$F_0^{B_c \chi_{c0}}$ & $ 0.47_{-0.05-0.01}^{+0.03+0.00}$  & $0.40_{-0.02-0.01}^{+0.00+0.00}$   & $-0.45_{-0.29-0.00}^{+0.33+0.00}$  & $-1.31_{-0.11-0.10}^{+0.12+0.09}$\\
 \hline\hline
 & $A^{B_c h_c}$ & $ 0.07_{-0.01-0.01}^{+0.00+0.01}$  & $0.11_{-0.01-0.01}^{+0.00+0.01}$   &  $2.32_{-0.26-0.11}^{+0.30+0.11}$ & $0.49_{-0.07-0.02}^{+0.07+0.01}$ \\
 & $V_0^{B_c h_c}$ & $0.64_{-0.00-0.02}^{+0.10+0.02}$  & $0.94_{-0.08-0.04}^{+0.03+0.03}$   & $1.92_{-0.32-0.05}^{+0.35+0.04}$  & $0.39_{-0.10-0.01}^{+0.08+0.00}$\\
 & $V_1^{B_ch_c}$ & $0.50_{-0.06-0.05}^{+0.03+0.04}$  & $0.65_{-0.05-0.05}^{+0.01+0.04}$   & $1.54_{-0.24-0.11}^{+0.29+0.11}$  & $0.24_{-0.06-0.02}^{+0.06+0.01}$ \\
 & $V_2^{B_ch_c}$ & $-0.32_{-0.04-0.03}^{+0.05+0.03}$  & $-0.55_{-0.03-0.05}^{+0.06+0.06}$   & $2.63_{-0.26-0.10}^{+0.28+0.09}$  & $0.63_{-0.07-0.01}^{+0.06+0.01}$\\
 \hline\hline
 & $A^{B_c\chi_{c1}}$ & $0.36_{-0.04-0.02}^{+0.02+0.01}$  & $0.54_{-0.03-0.04}^{+0.01+0.03}$   &  $1.98_{-0.23-0.11}^{+0.26+0.11}$ & $0.43_{-0.06-0.02}^{+0.04+0.01}$ \\
 & $V_0^{B_c  \chi_{c1}}$ & $0.13_{-0.01-0.01}^{+0.01+0.00}$  & $0.23_{-0.02-0.00}^{+0.01+0.00}$   & $2.99_{-0.13-0.29}^{+0.19+0.32}$  & $0.023_{-1.04-0.36}^{+0.32+0.16}$\\
 & $V_1^{B_c  \chi_{c1}}$ & $0.85_{-0.02-0.03}^{+0.00+0.02}$  & $0.73_{-0.06-0.04}^{+0.05+0.03}$   & $-0.51_{-0.35-0.10}^{+0.40+0.10}$  & $-1.38_{-0.17-0.02}^{+0.19+0.01}$ \\
 & $V_2^{B_c  \chi_{c1}}$ & $0.15_{-0.01-0.01}^{+0.01+0.01}$  & $0.19_{-0.01-0.02}^{+0.00+0.02}$   & $1.22_{-0.16-0.20}^{+0.18+0.20}$  & $-0.08_{-0.05-0.11}^{+0.26+0.09}$\\
 \hline\hline
 & $h^{B_c \chi_{c2}}$ & $0.022_{-0.003-0.001}^{+0.002+0.001}$  & $0.036_{-0.004-0.003}^{+0.002+0.003}$   &  $2.58_{-0.25-0.10}^{+0.28+0.01}$ & $0.61_{-0.06-0.01}^{+0.06+0.02}$ \\
 &$k^{B_c \chi_{c2}}$ & $1.27_{-0.13-0.07}^{+0.15+0.05}$  & $1.73_{-0.25-0.14}^{+0.12+0.11}$   & $1.61_{-0.21-0.12}^{+0.21+0.12}$  & $0.24_{-0.04-0.02}^{+0.01+0.02}$\\
 & $b_+^{B_c  \chi_{c2}}$ & $-0.011_{-0.000-0.000}^{+0.001+0.000}$  & $-0.018_{-0.001-0.000}^{+0.002+0.000}$   & $2.27_{-0.41-0.07}^{+0.36+0.07}$  & $0.46_{-0.25-0.01}^{+0.10+0.01}$ \\
 &$b_-^{B_c \chi_{c2}}$ & $0.020_{-0.00-0.00}^{+0.00+0.00}$  & $0.033_{-0.004-0.003}^{+0.002+0.003}$   & $2.48_{-0.24-0.11}^{+0.26+0.13}$  & $0.56_{-0.06-0.02}^{+0.05+0.11}$\\
 \hline\hline
\end{tabular}
\end{table}

\section{Semileptonic $B_c$ decays }\label{sec:semileptonic}

With the form factors at hand, one can directly perform the analysis
of semileptonic $B_c$ decays whose differential decay widths are
given by{\footnotesize
\begin{eqnarray}
 \frac{d\Gamma(B_c\to Sl\bar\nu)}{dq^2} &=&\left(\frac{q^2-m_l^2}{q^2}\right)^2\frac{ {\sqrt{\lambda(m_{B_c}^2,m_S^2,q^2)}} G_F^2 V_{cb}^2} {384m_{B_c}^3\pi^3}
 \times \frac{1}{q^2} \nonumber\\
 &&\;\;\;\times \left\{ (m_l^2+2q^2) \lambda(m_{B_c}^2,m_S^2,q^2) [F_1^{BcS}(q^2)]^2
 +3 m_l^2(m_{B_c}^2-m_S^2)^2 [F_0^{BcS}(q^2)]^2
 \right\},\\
 \frac{d\Gamma_L(B_c\to Al\bar\nu)}{dq^2}&=&\left(\frac{q^2-m_l^2}{q^2}\right)^2\frac{ {\sqrt{\lambda(m_{B_c}^2,m_A^2,q^2)}}
 G_F^2 V_{cb}^2} {384m_{B_c}^3\pi^3}
 \times \frac{1}{q^2} \left\{ 3 m_l^2 \lambda(m_{B_c}^2,m_A^2,q^2) [V_0^{BcA}(q^2)]^2\right.\nonumber\\
 && + \left.(m_l^2+2q^2) \left|\frac{1}{2m_A}  \left[
 (m_{B_c}^2-m_A^2-q^2)(m_{B_c}-m_A)V_1^{BcA}(q^2)-\frac{\lambda(m_{B_c}^2,m_A^2,q^2)}{m_{B_c}-m_A}V_2^{BcA}(q^2)\right]\right|^2
 \right\},\label{eq:BRBtoAL}\\
 \frac{d\Gamma^\pm(B_c\to Al\bar\nu)}{dq^2}&=&\left(\frac{q^2-m_l^2}{q^2}\right)^2\frac{
 {\sqrt{\lambda(m_{B_c}^2,m_A^2,q^2)}} G_F^2V_{cb}^2}{384m_{B_c}^3\pi^3}
 \nonumber\\
 && \times \left\{ (m_l^2+2q^2) \lambda(m_{B_c}^2,m_A^2,q^2)\left|\frac{A^{BcA}(q^2)}{m_{B_c}-m_A}\mp
 \frac{(m_{B_c}-m_A)V_1^{BcA}(q^2)}{\sqrt{\lambda(m_{B_c}^2,m_A^2,q^2)}}\right|^2
 \right\},
\end{eqnarray}}
where the superscript $+(-)$ denotes the right-handed (left-handed)
states of axial-vector mesons, while the subscript $L$ denotes that
the axial-vector in the final state is longitudinally polarized.
$m_l$ is the lepton's mass and $\lambda(m_{B_c}^2,
m_{i}^2,q^2)=(m_{B_c}^2+m_{i}^2-q^2)^2-4m_{B_c}^2m_i^2$ with
$i=S,A$.   The combined transverse and total differential decay
widths are given by:
\begin{eqnarray}
 \frac{d\Gamma_T}{dq^2}= \frac{d\Gamma^+}{dq^2}+
 \frac{d\Gamma^-}{dq^2},\;\;\;
 \frac{d\Gamma}{dq^2}= \frac{d\Gamma_L}{dq^2}+
 \frac{d\Gamma_T}{dq^2}.
\end{eqnarray}
Expressions for the decay width of $B_c\to Tl\bar\nu$ can be
obtained by the decay width of $B_c\to Al\bar\nu$ decays:
\begin{eqnarray}
 \frac{d\Gamma_L(B\to Tl\bar\nu)}{dq^2}&=&\frac{1}{2}\frac{\lambda(m_{B_c}^2,m_T^2,q^2)}{4m_T^2}\times
 \frac{d\Gamma_L(B\to
 Al\bar\nu)}{dq^2}|_{V_{0,1,2}^{B_cA}\to V_{0,1,2}^{B_cT}},\nonumber\\
 d\Gamma^{\pm}(B\to Tl\bar\nu)&=&\frac{2}{3}\frac{\lambda(m_{B_c}^2,m_T^2,q^2)}{4m_T^2}
 \times \frac{d\Gamma^{\pm}(B\to
 Al\bar\nu)}{dq^2}|_{(V_{1}^{B_cA},A^{B_cA})\to (V_{1}^{B_cT},A^{B_cT})},
\end{eqnarray}
where the form factors $V_{0,1,2},A$ of $B_c\to T$ decays have
nontrivial dimensions and thus the two functions
$\frac{d\Gamma_L(B\to
 Al\bar\nu)}{dq^2}|_{V_{0,1,2}^{B_c A}\to V_{0,1,2}^{B_cT}}$ and $\frac{d\Gamma^{\pm}(B\to
 Al\bar\nu)}{dq^2}|_{(V_{1}^{B_cA},A^{B_cA})\to (V_{1}^{B_cT},A^{B_cT})}$ do not have
the correct dimensions with the conventional $\frac{d\Gamma}{dq^2}$.
It is compensated by the prefactor
$\frac{\lambda(m_{B_c}^2,m_T^2,q^2)}{4m_T^2}$ which also have
nonzero mass dimensions.

Our results for these semileptonic $B_c$ decays are collected in
table~\ref{tab:BR}. Since electrons and muons are very light
compared with the charm quark, we can safely neglect the masses of
these two kinds of leptons. The uncertainties are from those in the
form factors: the first kind of uncertainties are from the shape
parameters of the charmonia  and the second ones are from the $B_c$
decay constant. The third uncertainties are from that in the $B_c$
lifetime.  Several remarks on the results are given in order. First
of all, branching fractions for the $B_c\to
(h_c,\chi_{c0,1,2})l\bar\nu (l=e,\mu)$ decays have the order
$10^{-3}$ while branching fractions of $B_c\to
(h_c,\chi_{c0,1,2})\tau\bar\nu_\tau$ are suppressed by one order. In
the covariant light-front quark model, branching fractions of the
$B_c\to \eta_c l\bar\nu$ and $B_c\to J/\psi l\bar\nu$ decays are
about one percent~\cite{Wang:2007sxa}:
\begin{eqnarray}
 {\cal BR}(B_c\to \eta_c e\bar\nu_e)=(0.67^{+0.04+0.04+0.10}_{-0.07-0.04-0.10})\%,\;\;\; {\cal BR}(B_c\to J/\psi
 e\bar\nu_e)=(1.49^{+0.01+0.15+0.23}_{-0.03-0.14-0.23})\%.
\end{eqnarray}
Compared with these decays, the branching ratios of $B_c\to
(h_c,\chi_{c0,1,2})l\bar\nu$ are smaller by a factor of $2-10$.
There are two main reasons for these differences: the form factors
and phase space. For example, if we set the mass of $\chi_{c0}$
equal to that of $\eta_c$, the branching ratio of $B_c\to
\chi_{c0}l\bar\nu$ becomes $0.40\%$. The larger form factors of
$B_c\to \eta_c$  will enhance the branching fraction by a factor of
1.68. Secondly,  polarizations $\frac{\Gamma_L}{\Gamma_T}$ of
$B_c\to h_cl\bar\nu$ and $B_c\to \chi_{c1}l\bar\nu$ are dramatically
different. As indicated from table~\ref{tab:resultsformfactors1},
the form factors $V_1$ and $V_2$ for $B_c\to h_c$ have different
signs. Thus the longitudinally polarized decay width receives
constructive contributions as we can see in Eq.~(\ref{eq:BRBtoAL}).
The form factor $A^{B_c h_c}$ is small which suppresses the
transversely polarized decay width. Accordingly,  a large
$\frac{\Gamma_L}{\Gamma_T}$ is expected:
$\frac{\Gamma_L}{\Gamma_T}\simeq11.1$ for $B_c\to h_c l\bar\nu$
decays and $\frac{\Gamma_L}{\Gamma_T}\simeq4.7$ for $B_c\to h_c
\tau\bar\nu_{\tau}$. The situation is dramatically different for
$B_c\to \chi_{c1}$ decays. The form factors $V_1^{B_c \chi_{c1}}$
and $V_2^{B_c\chi_{c1}}$ have the same sign, which gives destructive
contributions to the longitudinally polarized decay width. Form
factors $A^{B_c \chi_{c1}}$ and $V_1^{B_c \chi_{c1}}$ are large and
thus the minus polarized decay width is large. The polarization
fraction $\frac{\Gamma_L}{\Gamma_T}$ is reduced and predicted as:
$\frac{\Gamma_L}{\Gamma_T}\simeq0.24$ for $B_c\to \chi_{c1}
l\bar\nu$ decays and $\frac{\Gamma_L}{\Gamma_T}\simeq0.27$ for
$B_c\to \chi_{c1} \tau\bar\nu_{\tau}$. Compared with results in
Refs.~\cite{Chang:2001pm,Ivanov:2006ni,Ivanov:2005fd,Hernandez:2006gt}
for the $B_c\to (h_c,\chi_{c0,1,2})l\bar\nu$ decays which are
collected in table~\ref{tab:BR}, we can see that most of our
predictions on the semileptonic $B_c$ decays are comparable with
their predictions. These results will be tested at the ongoing and
forthcoming hadron colliders.

\begin{table}
\caption{Branching ratios  (in units of $\%$)    of    semileptonic
$B_c\to (h_c,\chi_{c0,1,2})l\bar\nu (l=e,\mu)$ and $B_c\to
(h_c,\chi_{c0,1,2}) \tau\bar\nu_\tau$ decays.   }\label{tab:BR}
\begin{tabular}{|c|c|c|c|c|c|c|c|c|c|c|}
 \hline\hline
                    & $B_c\to \chi_{c0}l\bar\nu$
                    & $B_c\to \chi_{c1}l\bar\nu$
                    & $B_c\to h_cl\bar\nu$
                    & $B_c\to \chi_{c2}l\bar\nu$  \\\hline
 This work & $0.21_{-0.04-0.01-0.03}^{+0.02+0.01+0.03}$  & $0.14_{-0.01-0.01-0.02}^{+0.00+0.01+0.02}$    & $0.31_{-0.08-0.01-0.05}^{+0.05+0.01+0.05}$   & $0.17_{-0.06-0.02-0.03}^{+0.04+0.02+0.03}$  \\\hline
 CCWZ~\cite{Chang:2001pm}& 0.12   &  0.15  & 0.18  & 0.19  \\\hline
 IKS~\cite{Ivanov:2006ni}& 0.17   &  0.092   & 0.27  & 0.17  \\\hline
 IKS~\cite{Ivanov:2005fd}& 0.18   &  0.098   & 0.31  & 0.20 \\\hline
 HNV\cite{Hernandez:2006gt}& 0.11   &  0.066    &0.17   & 0.13  \\
 \hline\hline
                     & $B_c\to \chi_{c0}\tau\bar\nu_{\tau}$
                     &$B_c\to \chi_{c1}\tau\bar\nu_{\tau}$
                     & $B_c\to h_c\tau\bar\nu_{\tau}$
                     & $B_c\to \chi_{c2}\tau\bar\nu_{\tau}$\\\hline
 This work   & $0.024_{-0.003-0.001-0.004}^{+0.001+0.001+0.004} $  &$0.015_{-0.001-0.002-0.002}^{+0.000+0.001+0.002}$   &$0.022_{-0.004-0.000-0.003}^{+0.002+0.000+0.003}$  & $0.0096_{-0.0029-0.0014-0.0015}^{+0.0019+0.0013+0.0015}$\\\hline
 CCWZ~\cite{Chang:2001pm}  & 0.017  & 0.024  &0.025    & 0.029\\\hline
 IKS~\cite{Ivanov:2006ni} & 0.013   & 0.0089  &0.017    & 0.0082\\\hline
 IKS~\cite{Ivanov:2005fd} & 0.018   & 0.012  &0.027   & 0.014\\\hline
 HNV\cite{Hernandez:2006gt}  & 0.013 & 0.0072   &0.015   & 0.0093\\
 \hline\hline
\end{tabular}
\end{table}

\section{Conclusion}\label{conc}

Due to its unique properties,  $B$ physics has attracted abundant
attentions. Measurements on the CKM matrix elements are becoming
more and more accurate, which makes the goal to test the CP origins
in and beyond SM much more practicable. $B_c$ meson decays provide
another promising place to continue the errand in $B$ meson decays,
and offer a new window to explore strong interactions. Although the
$B_c$ meson can not be generated on the two $B$ factories, it has a
promising prospect on the ongoing and forthcoming hadron colliders.
The high statistics of $B_c$ meson  at the forthcoming hadron
colliders can compensate for the hadronic pollution and make it
suitable for the precise determination of many standard model
parameters. Because of these interesting features, we have studied
the $B_c$ transition form factors in the covariant light-front quark
model, which are relevant for the semileptonic $B_c$ decays.

Branching fractions of $B_c\to (h_c,\chi_{c0,1,2})l\bar\nu
(l=e,\mu)$ decays have the order $10^{-3}$ while branching fractions
of $B_c\to (h_c,\chi_{c0,1,2})\tau\bar\nu_\tau$ are suppressed by
one order. Compared with  branching fractions of the $B_c\to \eta_c
l\bar\nu$ and $B_c\to J/\psi l\bar\nu$ decays,  the branching ratios
of $B_c\to (h_c,\chi_{c0,1,2})l\bar\nu$ are smaller by a factor of
$2-10$. The polarizations $\frac{\Gamma_L}{\Gamma_T}$ of $B_c\to
h_cl\bar\nu$ and $B_c\to \chi_{c1}l\bar\nu$ are dramatically
different: it is very large for $B_c\to h_cl\bar\nu$ but very small
for $B_c\to \chi_{c0}l\bar\nu$.  Most of our predictions on the
semileptonic $B_c$ decays are comparable with results in the
literature for the $B_c\to (h_c,\chi_{c0,1,2})l\bar\nu$ decays.
These results will be tested at the ongoing and forthcoming hadron
colliders.

\section*{Acknowledgement}

This work is partly supported by the National Natural Science
Foundation of China under the Grant No. 10735080, 10625525, and
10805037.

\appendix
%

\section{Some specific rules in the $p^-$ integration}\label{eqs:rules}

When performing the $p^-$ integration, we need to include the
zero-mode contribution. This amounts to performing the integration
in a proper way in this approach. To be more specific, for $\hat
p_1^\prime $ under integration we use the following
rules~\cite{Jaus:1999zv,Cheng:2003sm}
 \begin{eqnarray}
\hat p^\prime_{1\mu}
       &\doteq& P_\mu A_1^{(1)}+q_\mu A_2^{(1)},
\hat N_2
       \to Z_2,
  \nonumber\\
\hat p^\prime_{1\mu} \hat p^\prime_{1\nu}
       &\doteq& g_{\mu\nu} A_1^{(2)}+P_\mu P_\nu A_2^{(2)}+(P_\mu
                q_\nu+ q_\mu P_\nu) A^{(2)}_3+q_\mu q_\nu A^{(2)}_4,
                  \nonumber\\
\hat p^\prime_{1\mu} \hat p^\prime_{1\nu} \hat p^\prime_{1\alpha}
       &\doteq& (g_{\mu\nu} P_\alpha+g_{\mu\alpha} P_\nu+g_{\nu\alpha} P_\mu) A_1^{(3)}
               +(g_{\mu\nu} q_\alpha+g_{\mu\alpha} q_\nu+g_{\nu\alpha} q_\mu) A_2^{(3)}
  \nonumber \\
       &&       +P_\mu P_\nu P_\alpha A_3^{(3)}
                +(P_\mu P_\nu q_\alpha+ P_\mu q_\nu P_\alpha+q_\mu
                 P_\nu P_\alpha) A^{(3)}_4
  \nonumber \\
       &&       +(q_\mu q_\nu P_\alpha+ q_\mu P_\nu q_\alpha+P_\mu q_\nu q_\alpha)
                 A^{(3)}_5
                +q_\mu q_\nu q_\alpha  A^{(3)}_6,
   \nonumber\\
 \hat p_{1\mu}^\prime \hat N_2
        &\doteq& q_\mu\left[A^{(1)}_2 Z_2+\frac{q\cdot P}{q^2} A^{(2)}_1\right],
  \nonumber\\
\hat p_{1\mu}^\prime \hat p_{1\nu}^\prime \hat N_2
       &\doteq& g_{\mu \nu} A^{(2)}_1 Z_2+q_\mu q_\nu
             \left[A^{(2)}_4 Z_2+2\frac{q\cdot P}{q^2} A^{(1)}_2 A^{(2)}_1\right],
 \label{eq:p1B}
 \end{eqnarray}
where the symbol $\doteq$ reminds us that the above equations are
true only after integration. In the above equation, $A^{(i)}_j$ are
functions of $x_{1,2}$, $p^{\prime2}_\bot$, $p^\prime_\bot\cdot
q_\bot$ and $q^2$. Their explicit expressions have been studied in
Ref.~\cite{Jaus:1999zv,Cheng:2003sm}:
 \begin{eqnarray}
 Z_2&=&\hat N_1^\prime+m_1^{\prime2}-m_2^2+(1-2x_1)M^{\prime2}
 +(q^2+q\cdot P)\frac{p^\prime_\bot\cdot q_\bot}{q^2},  \nonumber\\
  A^{(1)}_1&=&\frac{x_1}{2},
 \quad
 A^{(1)}_2=A^{(1)}_1-\frac{p^\prime_\bot\cdot q_\bot}{q^2},
 \nonumber\\
 A^{(2)}_1&=&-p^{\prime2}_\bot-\frac{(p^\prime_\bot\cdot q_\bot)^2}{q^2},
 \quad
 A^{(2)}_2=\big(A^{(1)}_1\big)^2,
 \quad
 A^{(2)}_3=A^{(1)}_1 A^{(1)}_2,
 \nonumber\\
 A^{(2)}_4&=&\big(A^{(1)}_2\big)^2-\frac{1}{q^2}A^{(2)}_1,
 \quad
 A^{(3)}_1=A^{(1)}_1 A^{(2)}_1,
 \quad
 A^{(3)}_2=A^{(1)}_2 A^{(2)}_1,
 \nonumber\\
 A^{(3)}_3&=&A^{(1)}_1 A^{(2)}_2,
 \quad
 A^{(3)}_4=A^{(1)}_2 A^{(2)}_2,
 \quad
 A^{(3)}_5=A^{(1)}_1 A^{(2)}_4,
  \nonumber\\
 A^{(3)}_6&=&A^{(1)}_2 A^{(2)}_4-\frac{2}{q^2}A^{(1)}_2
 A^{(2)}_1.
 \label{eq:rule}
 \end{eqnarray}
We do not show the spurious contributions in Eq.~(\ref{eq:rule})
since they are numerically vanishing.

\section{Expressions of $B_c\to A,T$ form factors}
In this appendix, we collect the analytic expressions of $B_c\to
h_c,\chi_{1,2}$ form factors in the covariant light-front quark
model.
\begin{eqnarray}
A^{B_ch_c}(q^2)&=&(M^{\prime}-M^{\prime\prime})\frac{N_c}{16\pi^3}\int
dx_2 d^2 p^\prime_\bot
           \frac{ 2h^\prime_P h^\pp_{^1A}}{x_2 \hat N^\prime_1 \hat N^\pp_1}
             \frac{1}{w^\pp_{^1A}}p^{\prime2}_\bot,
        \\
 V_1^{B_ch_c}(q^2)&=&\frac{1}{M^{\prime}-M^{\prime\prime}}\frac{N_c}{16\pi^3}\int dx_2 d^2 p^\prime_\bot
            \frac{ h^\prime_P h^\pp_{^1A}}{x_2 \hat N^\prime_1 \hat N^\pp_1}
             \bigg\{4\frac{q^2 p^{\prime2}_\bot+(p^\prime_\bot\cdot q_\bot)^2}{q^2
w^\pp_{^1A}}
            \bigg[2 x_1 (M^{\prime2}+M^{\prime2}_0)
 \nonumber\\
         &&-q^2-q\cdot P-2(q^2+q\cdot P)\frac{p^\prime_\bot\cdot
            q_\bot}{q^2}-2(m_1^\prime+m_1^\pp)(m_1^\prime-m_2)
            \bigg]\bigg\},
   \\
 V_2^{B_ch_c}(q^2)&=&-(M^{\prime}-M^{\prime\prime})\frac{N_c}{16\pi^3}\int dx_2 d^2 p^\prime_\bot
            \frac{2 h^\prime_P h^\pp_{^1A}}{x_2 \hat N^\prime_1 \hat N^\pp_1}
            \nonumber\\
         && 2\frac{x_2 q^2+p_\bot^\prime\cdot q_\bot}{x_2 q^2
            w^\pp_{^1A}}\Big[p^\prime_\bot\cdot p^\pp_\bot+(x_1 m_2+x_2 m_1^\prime)(x_1 m_2+x_2
            m_1^\pp)\Big],
   \\
 V_0^{B_ch_c}(q^2)&=&\frac{M^{\prime}-M^\pp}{2M^\pp}V_1^{B_ch_c}(q^2)
 -\frac{M^{\prime}+M^\pp}{2M^\pp}V_2^{B_ch_c}(q^2)-\frac{q^2}{2M^\pp}
         \frac{N_c}{16\pi^3}\int dx_2 d^2 p^\prime_\bot
            \frac{ h^\prime_P h^\pp_{^1A}}{x_2 \hat N^\prime_1 \hat N^\pp_1}
 \nonumber\\
         &&\frac{4}{w^\pp_{^1A}}\bigg([M^{\prime2}+M^{\pp2}-q^2+2(m_1^\prime-m_2)(m_2-m_1^\pp)]
                                   (A^{(2)}_3+A^{(2)}_4-A^{(1)}_2)
 \nonumber\\
         &&+Z_2(3 A^{(1)}_2-2A^{(2)}_4-1)+\frac{1}{2}[x_1(q^2+q\cdot P)
            -2 M^{\prime2}-2 p^\prime_\bot\cdot q_\bot -2 m_1^\prime(m_2-m_1^\pp)
 \nonumber\\
         &&-2 m_2(m_1^\prime-m_2)](A^{(1)}_1+A^{(1)}_2-1)+
         q\cdot P\Bigg[\frac{p^{\prime2}_\bot}{q^2} +\frac{(p^\prime_\bot\cdot q_\bot)^2}{q^4}\Bigg]
         (4A^{(1)}_2-3)\bigg).
 \end{eqnarray}
\begin{eqnarray}
 A^{B_c\chi_{c1}}(q^2)&=&(M^{\prime}-M^{\prime\prime})\frac{N_c}{16\pi^3}\int dx_2 d^2 p^\prime_\bot
           \frac{2 h^\prime_P h^\pp_{^3A}}{x_2 \hat N^\prime_1 \hat N^\pp_1}
           \Bigg\{x_2 m_1^\prime+x_1 m_2+(m_1^\prime+m_1^\pp)
           \frac{p^\prime_\bot\cdot q_\bot}{q^2}
            \Bigg\},
   \\
  V_1^{B_c\chi_{c1}}(q^2)&=&-\frac{1}{M^{\prime}-M^{\prime\prime}}\frac{N_c}{16\pi^3}\int dx_2 d^2 p^\prime_\bot
            \frac{ h^\prime_P h^\pp_{^3A}}{x_2 \hat N^\prime_1 \hat N^\pp_1}
            \Bigg\{2
            x_1(m_2-m_1^\prime)(M^{\prime2}_0+M^{\pp2}_0)+4 x_1
            m_1^\pp M^{\prime2}_0
  \nonumber\\
         &&+2x_2 m_1^\prime q\cdot P+2 m_2 q^2-2 x_1 m_2
           (M^{\prime2}+M^{\pp2})+2(m_1^\prime-m_2)(m_1^\prime-m_1^\pp)^2
  \nonumber\\
         &&
           +8(m_1^\prime-m_2)\left[p^{\prime2}_\bot+\frac{(p^\prime_\bot\cdot
            q_\bot)^2}{q^2}\right]+2(m_1^\prime-m_1^\pp)(q^2+q\cdot
           P)\frac{p^\prime_\bot\cdot q_\bot}{q^2}
           \Bigg\},
   \\
 V_2^{B_c\chi_{c1}}(q^2)&=&(M^{\prime}-M^{\prime\prime})\frac{N_c}{16\pi^3}\int dx_2 d^2 p^\prime_\bot
            \frac{2 h^\prime_P h^\pp_{^3A}}{x_2 \hat N^\prime_1 \hat N^\pp_1}
            \nonumber\\
         &&\Bigg\{(x_1-x_2)(x_2 m_1^\prime+x_1 m_2)-[2x_1
            m_2-m_1^\pp+(x_2-x_1)
            m_1^\prime]\frac{p^\prime_\bot\cdot q_\bot}{q^2}\Bigg\},
   \\
  V_0^{B_c\chi_{c1}}(q^2)&=&\frac{M^{\prime}-M^\pp}{2M^\pp}V_1^{B_c\chi_{c1}}(q^2)
  -\frac{M^{\prime}+M^\pp}{2M^\pp}V_2^{B_c\chi_{c1}}(q^2)
  -\frac{q^2}{2M^\pp}
 \frac{N_c}{16\pi^3}\int dx_2 d^2 p^\prime_\bot
            \frac{ h^\prime_P h^\pp_{^3A}}{x_2 \hat N^\prime_1 \hat N^\pp_1}
            \nonumber\\
         &&\Bigg\{2(2x_1-3)(x_2 m_1^\prime+x_1 m_2)-8(m_1^\prime-m_2)
            \left[\frac{p^{\prime2}_\bot}{q^2}+2\frac{(p^\prime_\bot\cdot q_\bot)^2}{q^4}\right]
 \nonumber\\
         &&-[(14-12 x_1) m_1^\prime+2 m_1^\pp-(8-12 x_1) m_2]\frac{p^\prime_\bot\cdot q_\bot}{q^2}
            \Bigg\}.
 \end{eqnarray}
Notice that $m^\pp_1=m_2=m_c$ in the case of the outgoing p-wave
charmonia, so $\frac{1}{w '' _{^3A}}$ is zero, leading to the
vanishing term of $\frac{1}{w '' _{^3A}}$ in the form factors.
\begin{eqnarray}
 h^{B_c\chi_{c2}}(q^2)&=&\frac{N_c}{16\pi^3}\int dx_2 d^2 p^\prime_\bot
           \frac{2 h^\prime_P h^\pp_T}{x_2 \hat N^\prime_1 \hat N^\pp_1}
           \Bigg\{x_2 m_1^\prime+x_1 m_2+(m_1^\prime-m_1^\pp)
           \frac{p^\prime_\bot\cdot q_\bot}{q^2}
\nonumber\\
          && +\frac{2}{w^\pp_T}\left[p^{\prime2}_\bot+\frac{(p^\prime_\bot\cdot
            q_\bot)^2}{q^2}\right]+\bigg[(m^\prime_1-m^\pp_1)(A^{(2)}_3+A^{(2)}_4)+
           (m^\pp_1+m^\prime_1-2m_2)(A^{(2)}_2+A^{(2)}_3)
\nonumber\\
          &&-m^\prime_1(A^{(1)}_1+A^{(1)}_2)+\frac{2}{w^\pp_T}(2A^{(3)}_1+2A^{(3)}_2-A^{(2)}_1)\bigg]
           \Bigg\},
  \\
 k^{B_c\chi_{c2}}(q^2)&=&-\frac{N_c}{16\pi^3}\int dx_2 d^2 p^\prime_\bot
            \frac{ h^\prime_P h^\pp_T}{x_2 \hat N^\prime_1 \hat N^\pp_1}
            \Bigg\{2
            x_1(m_2-m_1^\prime)(M^{\prime2}_0+M^{\pp2}_0)-4 x_1
            m_1^\pp M^{\prime2}_0+2x_2 m_1^\prime q\cdot P
  \nonumber\\
         &&+2 m_2 q^2-2 x_1 m_2
           (M^{\prime2}+M^{\pp2})+2(m_1^\prime-m_2)(m_1^\prime+m_1^\pp)^2
  \nonumber\\
         &&+8(m_1^\prime-m_2)\left[p^{\prime2}_\bot+\frac{(p^\prime_\bot\cdot
            q_\bot)^2}{q^2}\right]
           +2(m_1^\prime+m_1^\pp)(q^2+q\cdot
           P)\frac{p^\prime_\bot\cdot q_\bot}{q^2}
 \nonumber\\
         &&-4\frac{q^2 p^{\prime2}_\bot+(p^\prime_\bot\cdot q_\bot)^2}{q^2 w^\pp_T}
            \bigg[2 x_1 (M^{\prime2}+M^{\prime2}_0)-q^2-q\cdot P-2(q^2+q\cdot P)\frac{p^\prime_\bot\cdot
            q_\bot}{q^2}
 \nonumber\\
         &&-2(m_1^\prime-m_1^\pp)(m_1^\prime-m_2)
            \bigg]\Bigg\}
  \nonumber\\
         &&+\frac{N_c}{16\pi^3}\int dx_2 d^2 p^\prime_\bot
            \frac{ h^\prime_P h^\pp_T}{x_2 \hat N^\prime_1 \hat N^\pp_1}\Bigg\{2(A^{(1)}_1+A^{(1)}_2)[m_2(q^2-\hat N^\prime_1-\hat N^\pp_1-m^{\prime2}_1-m^{\pp2}_1)
 \nonumber\\
         &&-m^\prime_1(M^{\pp2}-\hat N^\pp_1-m^{\pp2}_1-m^2_2)-m^\pp_1(M^{\prime2}-\hat N^\prime_1-m^{\prime2}_1-m^2_2)-2m^\prime_1m^\pp_1m_2]
  \nonumber\\
         &&+2(m^\prime_1+m^\pp_1)(A^{(1)}_2Z_2+\frac{P\cdot q}{q^2}A^{(2)}_1)+16(m_2-m^\prime_1)(A^{(3)}_1+A^{(3)}_2)+4(2m^\prime_1-m^\pp_1-m_2)A^{(2)}_1
  \nonumber\\
         &&+\frac{4}{w^\pp_T}\bigg([M^{\prime2}+M^{\pp2}-q^2+2(m^\prime_1-m_2)(m^\pp_1+m_2)](2A^{(3)}_1+2A^{(3)}_2-A^{(2)}_1)
   \nonumber\\
         &&-4[A^{(3)}_2Z_2+\frac{P\cdot q}{3q^2}(A^{(2)}_1)^2]+2A^{(2)}_1Z_2\bigg)\Bigg\},
  \\
 b_+^{B_c\chi_{c2}}(q^2)&=&-\frac{N_c}{16\pi^3}\int dx_2 d^2 p^\prime_\bot
            \frac{2 h^\prime_P h^\pp_T}{x_2 \hat N^\prime_1 \hat N^\pp_1}
            \Bigg\{(x_1-x_2)(x_2 m_1^\prime+x_1 m_2)-[2x_1
            m_2+m_1^\pp+(x_2-x_1)
            m_1^\prime]\frac{p^\prime_\bot\cdot q_\bot}{q^2}
  \nonumber\\
         &&-2\frac{x_2 q^2+p_\bot^\prime\cdot q_\bot}{x_2 q^2
            w^\pp_T}\Big[p^\prime_\bot\cdot p^\pp_\bot+(x_1 m_2+x_2 m_1^\prime)(x_1 m_2-x_2
            m_1^\pp)\Big]\Bigg\}
  \nonumber\\
         &&+\frac{N_c}{16\pi^3}\int dx_2 d^2 p^\prime_\bot
            \frac{h^\prime_P h^\pp_T}{x_2 \hat N^\prime_1 \hat N^\pp_1}
            \Bigg\{8(m_2-m^\prime_1)(A^{(3)}_3+2A^{(3)}_4+A^{(3)}_5)-2m^\prime_1(A^{(1)}_1+A^{(1)}_2)
  \nonumber\\
         &&+4(2m^\prime_1-m^\pp_1-m_2)(A^{(2)}_2+A^{(2)}_3)+2(m^\prime_1+m^\pp_1)(A^{(2)}_2+2A^{(2)}_3+A^{(2)}_4)
  \nonumber\\
         &&+\frac{2}{w^\pp_T}\bigg[2[M^{\prime2}+M^{\pp2}-q^2+2(m^\prime_1-m_2)(m^\pp_1+m_2)](A^{(3)}_3+2A^{(3)}_4+A^{(3)}_5-A^{(2)}_2-A^{(2)}_3)
  \nonumber\\
         &&+[q^2-\hat N^\prime_1-\hat N^\pp_1-(m^\prime_1+m^\pp_1)^2](A^{(2)}_2+2A^{(2)}_3+A^{(2)}_4-A^{(1)}_1-A^{(1)}_2)\bigg]
            \Bigg\},
 \\
 b_-^{B_c\chi_{c2}}(q^2)&=&-\frac{N_c}{16\pi^3}\int dx_2 d^2 p^\prime_\bot
            \frac{ h^\prime_P h^\pp_T}{x_2 \hat N^\prime_1 \hat N^\pp_1}
            \Bigg\{2(2x_1-3)(x_2 m_1^\prime+x_1 m_2)-8(m_1^\prime-m_2)
            \left[\frac{p^{\prime2}_\bot}{q^2}+2\frac{(p^\prime_\bot\cdot q_\bot)^2}{q^4}\right]
 \nonumber\\
         &&-[(14-12 x_1) m_1^\prime-2 m_1^\pp-(8-12 x_1) m_2]\frac{p^\prime_\bot\cdot q_\bot}{q^2}\nonumber\\
         &&+\frac{4}{w^\pp_T}\bigg([M^{\prime2}+M^{\pp2}-q^2+2(m_1^\prime-m_2)(m_1^\pp+m_2)]
                                   (A^{(2)}_3+A^{(2)}_4-A^{(1)}_2)
 \nonumber\\
         &&+Z_2(3 A^{(1)}_2-2A^{(2)}_4-1)+\frac{1}{2}[x_1(q^2+q\cdot P)
            -2 M^{\prime2}-2 p^\prime_\bot\cdot q_\bot -2 m_1^\prime(m_1^\pp+m_2)
 \nonumber\\
         &&-2 m_2(m_1^\prime-m_2)](A^{(1)}_1+A^{(1)}_2-1)+
         q\cdot P\Bigg[\frac{p^{\prime2}_\bot}{q^2} +\frac{(p^\prime_\bot\cdot q_\bot)^2}{q^4}\Bigg] (4A^{(1)}_2-3)\bigg)
            \Bigg\}
\nonumber\\
         &&+\frac{N_c}{16\pi^3}\int dx_2 d^2 p^\prime_\bot
            \frac{h^\prime_P h^\pp_T}{x_2 \hat N^\prime_1 \hat N^\pp_1}
            \Bigg\{8(m_2-m^\prime_1)(A^{(3)}_4+2A^{(3)}_5+A^{(3)}_6)-6m^\prime_1(A^{(1)}_1+A^{(1)}_2)
  \nonumber\\
         &&+4(2m^\prime_1-m^\pp_1-m_2)(A^{(2)}_3+A^{(2)}_4)+2(3m^\prime_1+m^\pp_1-2m_2)(A^{(2)}_2+2A^{(2)}_3+A^{(2)}_4)
  \nonumber\\
         &&+\frac{2}{w^\pp_T}\bigg[2[M^{\prime2}+M^{\pp2}-q^2+2(m^\prime_1-m_2)(m^\pp_1+m_2)](A^{(3)}_4+2A^{(3)}_5+A^{(3)}_6-A^{(2)}_3-A^{(2)}_4)
 \nonumber\\
         &&+2Z_2(3A^{(2)}_4-2A^{(3)}_6-A^{(1)}_2)+2\frac{q\cdot
         P}{q^2}(6A^{(1)}_2A^{(2)}_1-6A^{(1)}_2A^{(3)}_2+\frac{2}{q^2}(A^{(2)}_1)^2-A^{(2)}_1)
  \nonumber\\
         &&+[q^2-2M^{\prime2}+\hat N^\prime_1-\hat
         N^\pp_1-(m^\pp_1+m_2)^2+2(m^\prime_1-m_2)^2](A^{(2)}_2+2A^{(2)}_3+A^{(2)}_4-A^{(1)}_1-A^{(1)}_2).
 \end{eqnarray}

\end{document}